\documentclass[aps,preprint,groupedaddress,showpacs,floatfix]{revtex4-1}
\usepackage{graphicx}
\begin{document}

\title{To the nature of $\chi_{c2}(2P)$: the two-gluon decay}

\author{N.N. Achasov}\email{achasov@math.nsc.ru}
\affiliation{Laboratory of Theoretical Physics,
 Sobolev Institute for Mathematics, 630090, Novosibirsk, Russia}

\author{Xian-Wei Kang}\email{kangxianwei1@gmail.com}
\affiliation{Institute of Physics, Academia Sinica, Taipei, Taiwan
115}

\date{\today}

\begin{abstract}
We expect that $BR(\chi_{c2}(2P)\to gluon\, gluon)\gtrsim 2 \%$ if
Particle Data Group as well as BaBar and Belle collaboration
correctly identified the state. In reality, this branching ratio
corresponds to the one for $\chi_{c2}(2P)$ decaying to the light
hadrons. We also discuss the detection possibilities of these
decays.
\end{abstract}

\maketitle

More and more $XYZ$ states are observed in experiment. The
interpretation on them is still challenging to the community. But
some of them may be just the quarkonium states.

In 2006, Belle collaboration reported a resonance with $5.3\sigma$
statistical significance of the signal via the $\gamma\gamma\to
D\bar D$ process \cite{Belle}. The properties of the mass, angular
distributions, and $\Gamma_{\gamma\gamma}\Gamma_{D\bar D}/\Gamma$
(see also the Eq.~(\ref{chic2(2p)BR2gamma}) below) are all
consistent with the $2\,{}^3P_2$ charmonium state, as now identified
by Particle Data Group (PDG) as $\chi_{c2}(2P)$ \cite{PDG17}. Later
BaBar collaboration \cite{BABAR} confirmed such observation using
the data samples of the comparable magnitude with Belle. Its mass
and width read $M=3927.2\pm2.6$ MeV and $\Gamma=24\pm6$ MeV,
respectively \cite{PDG17}. To date, only the quantity
$\Gamma_{\gamma\gamma}\Gamma_{D\bar D}/\Gamma$ has been determined,
except for the constrained upper limit for the product of the
branching ratio for $\gamma\gamma$ and some selected hadronic states
\cite{11M,12AE,13}. Little is known beyond them, and especially for
the branching ratio of hadronic decays. In this Letter, we will
predict the branching ratio of the decay $\chi_{c2}\to 2g$ ($g$
denoting gluon) in a reliable way, by combining the observation of
the known experimental facts and the successful application of the
charmonium model \cite{ITEP}. That branching ratio corresponds to
summing over the ones for light hadronic decay in the practice.

It is natural that in the  nonrelativistic potential model of
charmonium, the ratio of the two-photon and two-gluon widths of the
charmonium decays does not depend on the wave function and slowly
grows with increase of the charmonium mass because of the
proportionality to $1/\alpha_s^2$, see for example, Ref.
\cite{ITEP}. The well established states \cite{PDG17} confirm this
consideration:
\begin{eqnarray}
& BR(\eta_c(1S)\to 2\gamma)=\frac{\Gamma(\eta_c(1S)\to
2\gamma)}{\Gamma(\eta_c(1S))} =\frac{\Gamma(\eta_c(1S)\to
2\gamma)}{\Gamma(\eta_c(1S)\to 2g)}\approx 1.59\times
10^{-4}\,,\nonumber\\[12pt]
 &\frac{\Gamma(\chi_{c0}(1P)\to 2\gamma)}{\Gamma(\chi_{c0}(1P)\to 2g)}=
\frac{\Gamma(\chi_{c0}(1P)\to 2\gamma
)}{\Gamma(\chi_{c0}(1P))-\Gamma(\chi_{c0}(1P)\to\gamma
J/\psi(1S))} =\frac{BR(\chi_{c0}(1P)\to
2\gamma)}{1-BR(\chi_{c0}(1P)\to\gamma J/\psi(1S))}\approx
2.26\times 10^{-4}\,,\nonumber\\[12pt]
 &\frac{\Gamma(\chi_{c2}(1P)\to 2\gamma)}{\Gamma(\chi_{c2}(1P)\to 2g)}=
\frac{\Gamma(\chi_{c2}(1P)\to 2\gamma
)}{\Gamma(\chi_{c2}(1P))-\Gamma(\chi_{c2}(1P)\to\gamma J/\psi(1S))}
=\frac{BR(\chi_{c2}(1P)\to 2\gamma)}{1-BR(\chi_{c2}(1P)\to\gamma
J/\psi(1S))}\approx 3.39\times 10^{-4}\,,
 \label{known}
\end{eqnarray}
where we have used \cite{PDG17}
\begin{eqnarray}
BR(\chi_{c0}(1P)\to \gamma J/\psi(1S))=(1.27\pm0.06)\%\,,\nonumber\\
BR(\chi_{c2}(1P)\to \gamma J/\psi(1S))=(19.2\pm0.7)\%\,.
\end{eqnarray}
Note that according to QCD, the decay of charmonium is due to the
annihilation of $c\bar c$ pair. The mass of $c\bar c$ is large and
$c\bar c\to \text{gluons}$ are perturbative, so two-gluon decay mode
is dominant. In the above equations, we did not use $\eta_c(2S)$ as
argument. Its hadronic decay channels are not well determined yet,
and also the only one measured radiative channel $\eta_c(2S)\to
\gamma\gamma$ suffers from very large uncertainty.

We know that
\begin{eqnarray}
&& \Gamma(\chi_{c2}(2P)\to\gamma\gamma)BR(\chi_{c2}(2P)\to D\bar D)
=(0.24\pm0.05\pm0.04 )\,\mbox{keV}\ \ \mbox{\cite{BABAR}}\nonumber\\
&&\Gamma(\chi_{c2}(2P)\to\gamma\gamma)BR(\chi_{c2}(2P)\to D\bar D) =
 (0.18\pm0.05\pm0.03)\,\mbox{keV}\ \ \mbox{\cite{Belle}}.
\label{chic2(2p)}
\end{eqnarray}
The PDG average gives $0.21\pm0.04$ keV \cite{PDG17}.

Taking into account $\Gamma(\chi_{c2}(2P))\approx 24$ MeV
\cite{PDG17}, we find
\begin{eqnarray}
& BR(\chi_{c2}(2P)\to 2\gamma)BR(\chi_{c2}(2P)\to D\bar D)\approx
10^{-5}\ \ \mbox{or}\nonumber\\
 &
BR(\chi_{c2}(2P)\to 2\gamma)BR(\chi_{c2}(2P)\to D\bar D)\approx
0.75\times 10^{-5}.
 \label{chic2(2p)BR2gamma}
\end{eqnarray}

Conservatively selecting from Eq. (\ref{known}) the ratio of the
two-gluon to two-photon widths of the charmonium decays equals
around $(1/4)\times 10^{4}$, we obtain
\begin{eqnarray}
&& BR(\chi_{c2}(2P)\to 2g)BR(\chi_{c2}(2P)\to D\bar D)\approx
0.025\ \ \mbox{or}\nonumber\\
&& BR(\chi_{c2}(2P)\to 2g)BR(\chi_{c2}(2P)\to D\bar D)\approx 0.019.
 \label{chic2(2p)BR2g}
\end{eqnarray}
So we expect that $BR(\chi_{c2}(2P)\to 2g)\gtrsim (2\pm0.4)\%$ if
PDG correctly identified the state.

It is obvious that the hadron channels of the two-gluon decays of
$\chi_{c2}(2P)$ could be the same as in the $\chi_{c2}(1P)$ case,
that is, there are a few  tens of
 such channels. It is expected that the
difference in the radial wave functions of $\chi_{c2}(1P)$ and
$\chi_{c2}(2P)$ does not lead to a significant difference in
$\Gamma(\chi_{c2}(1P)\to\gamma\gamma)$ and
$\Gamma(\chi_{c2}(2P)\to\gamma\gamma)$. Indeed,
$\Gamma(\chi_{c2}(1P)\to\gamma\gamma)\approx 0.5$ keV \cite{PDG17}
and $\Gamma(\chi_{c2}(2P)\to\gamma\gamma)\gtrsim 0.24$ keV or 0.18
keV, cf. Eq.~(\ref{chic2(2p)}). That is to say, it is possible that
$\Gamma(\chi_{c2}(2P)\to\gamma\gamma)\approx 0.5$ keV because of the
$D\bar D^*+\bar D D^*$ channel which can be essential. For example,
assuming $BR(\chi_{c2}(2P)\to D\bar D)\approx BR(\chi_{c2}(2P)\to
D\bar D^*+\bar D D^*)$, then
$\Gamma(\chi_{c2}(2P)\to\gamma\gamma)\gtrsim 0.48$ keV or 0.36 keV.
It is also clear that such a consideration takes place for
$\Gamma(\chi_{c2}(1P)\to 2g)$ and $\Gamma(\chi_{c2}(2P)\to 2g)$ that
results in $\Gamma(\chi_{c2}(2P)\to 2g)\approx
\Gamma(\chi_{c2}(1P)\to 2g) =\Gamma
(\chi_{c2}(1P))(1-BR(\chi_{c2}(1P)\to\gamma J/\psi(1S))\approx 1.56$
MeV. We then obtain $BR(\chi_{c2}(2P)\to 2g)\approx 6.5 \% $. In
fact, the mass difference for $\chi_{c2}(2P)$ and $\chi_{c2}(1P)$
can be considered. In Ref.~\cite{Rosner} the heavy quark mass $m_Q$
is used in the non-relativistic limit, and instead, the meson mass
$M$ is adopted in Ref.~\cite{ITEP}, which leads to the difference of
30\%. Guided by this estimate, we will write
$\Gamma(\chi_{c2}(2P)\to2\gamma)\approx (0.5\pm0.2)$ keV, and
$BR(\chi_{c2}(2P)\to2 g)\approx (6.5\pm 2.0)\%$. The current
measurements have the uncertainties with the similar size. Then we
note that our such observations and results agree very well with an
explicit calculation from a heavy quark potential derived from the
instanton vacuum along with the Coulomb and linear confinement
potential \cite{IndiaPaper}.

The confirmation of $\chi_{c2}(2P)$ state can be tested by BESIII,
for example, through the process $e^+e^-\to\psi(4040)\to \gamma
\chi_{c2}(2P)$ using their data above the center of mass of 4 GeV,
in which detector the two $D$'s can be clearly reconstructed. The
search for the two-gluon decays of the $\chi_{c2}(2P)$ state is
feasible for BESIII as well as other super factories: the BaBar and
Belle collaborations.

The author XWK thanks stimulating discussion with Prof.~Xiang Liu.
He also acknowledges helpful discussion from Sadaharu Uehara for
Belle measurements, and from Hai-Bo Li for BESIII measurements. N.N.
Achasov was supported in part by RFBR, Grant No. 16-02-00065, and by
Presidium of the Russian Academy of Sciences, Project No.
0314-2015-0011. XWK's work is supported by MOST, Taiwan, under Grant
No. 104-2112-M-001-022.


\begin{thebibliography}{99}

\bibitem{Belle} S. Uehara et al. (Belle Callaboration), Phys.\
Rev.\ Lett.\ {\bf 99}, 082003 (2006).

\bibitem{PDG17} C. Patrignani et al.
(Particle Data Group), Chin.\ Phys.\ C {\bf 40}, 100001 (2016) and
2017 update.

\bibitem{BABAR} B. Aubert et al. (BABAR Callaboration), Phys.\
Rev.\ D {\bf 81}, 092003 (2010).

\bibitem{11M}
 P.~del Amo Sanchez {\it et al.} [BaBar Collaboration],
  Phys.\ Rev.\ D {\bf 84}, 012004 (2011).

\bibitem{12AE}
 J.~P.~Lees {\it et al.} [BaBar Collaboration],
  Phys.\ Rev.\ D {\bf 86}, 092005 (2012).

\bibitem{13}
 S.~Uehara {\it et al.} [Belle Collaboration],
  PTEP {\bf 2013}, no. 12, 123C01 (2013).

\bibitem{ITEP} V.~A.~Novikov, L.~B.~Okun, M.~A.~Shifman, A.~I.~Vainshtein,
M.~B.~Voloshin and V.~I.~Zakharov,
  Phys.\ Rept.\  {\bf 41}, 1 (1978).

\bibitem{Rosner}
 W.~Kwong, P.~B.~Mackenzie, R.~Rosenfeld and J.~L.~Rosner,
  Phys.\ Rev.\ D {\bf 37}, 3210 (1988).

\bibitem{IndiaPaper}
 P.~P.~D'Souza, M.~Bhat, A.~P.~Monteiro and K.~B.~Vijaya Kumar,
  arXiv:1703.10413 [hep-ph].

\end{thebibliography}
\end{document}